\documentclass[twocolumn,amsmath,aps]{revtex4}

\usepackage{epsfig}
\usepackage{subfigure}

\newcommand{\be}{\begin{equation}}
\newcommand{\ee}{\end{equation}}
\newcommand{\bea}{\begin{eqnarray}}
\newcommand{\eea}{\end{eqnarray}}

\begin{document}

\title{Errors in Estimating $\Omega_{\Lambda}$ due to the Fluid Approximation}
\author{Timothy Clifton}
\email{tclifton@astro.ox.ac.uk}
\affiliation{Department of Astrophysics, University of Oxford, UK}
\author{Pedro G. Ferreira}
\email{pgf@astro.ox.ac.uk}
\affiliation{Department of Astrophysics, University of Oxford, UK}

% ----------------------- ABSTRACT -------------------------

\begin{abstract}

The matter content of the Universe is strongly inhomogeneous on small scales.
Motivated by this fact, we consider a model of the Universe that has regularly spaced discrete
masses, rather than a continuous fluid.  The optical properties of such space-times can differ
considerably from the continuous fluid case, even if the `average'
dynamics are the same.  We show that these differences have
consequences for cosmological parameter estimation, and that fitting to recent
supernovae observations gives a correction to the inferred value of
$\Omega_{\Lambda}$ of $\sim 10\%$.

\end{abstract}

\maketitle

In the standard approach to relativistic cosmology it is usual to assume that
the cosmological principle implies that the Universe is 
permeated by a set of continuous perfect fluids. Examples of these are
photons, baryons, dark matter and dark energy. Yet when we observe
the Universe, there is clear evidence for discreteness. Matter is accumulated in
stars and galaxies which arrange themselves in clusters, filaments and walls. 
These structures occupy a small volume of space, the rest of which is
almost completely devoid of electromagnetically interacting matter. It
is therefore reasonable to ask if the fluid approximation is a good
representation of the real Universe, and, if not, what are the corrections
due to the discretization we observe.

In  \cite{1} (henceforth, paper I) we considered just such a model of
the Universe: The matter content was taken to be discrete
islands of mass, rather than the usual continuous fluid.
This approach, which we dubbed `Archipelagian Cosmology', was based on
the Lattice Universe model of Lindquist and Wheeler \cite{LW}, a construction analogous 
to that of Wigner and Seitz in electromagnetism \cite{WS}.  The principal result of the study
was that even if the large-scale dynamics of a universe with discretized matter
content approach those of a Friedmann-Robertson-Walker (FRW) universe,
this does not mean that the optical properties will. Observers in a
universe filled with discrete objects can measure different redshifts and
luminosity distances to distant astrophysical objects, even if the
`average' expansion is identical to FRW.

Of course, paper I is not the first study of the optical properties
of an inhomogeneous universe.  Previous attempts have been made in the
context of Swiss Cheese cosmologies \cite{Kant,bmn,bn,btt1,m,cz},
as well as with linear perturbations to FRW geometry
\cite{ref1,ref2,ref3,ref4,Marra}.  The difference in paper I is that
the model used does not rely on an FRW background (neither as an
embedding structure for the inhomogeneities, nor as a background to be
perturbed around).  Instead, the model is a bottom up construction
that is not FRW at any point in space-time, yet behaves dynamically like
FRW on large scales.  The model does not, therefore, rely on any ill-defined concepts such as `average'
geometry, and is free from ambiguities associated with the scale
of smoothing.  Hence, we expect this approach to provide us
with new insights, and, in particular, new ways
to test concepts such as the fluid approximation without the constraint
of being tied to FRW geometry.

In paper I we considered the simple case of Milky Way-sized spherically symmetric
masses arranged on a regular lattice with a spacing of $\sim 1$Mpc,
and with critical density. We found that in the absence of a Cosmological Constant the
deceleration parameter measured by an observer in such a space-time would
be $q_0 \simeq 8/7$, rather than the usual value of $1/2$ that would
be measured in a universe filled with a perfect fluid of dust (i.e. an
Einstein-de Sitter universe).  This result is due to two separate
effects.  Firstly, the null geodesics that connect observers and
sources pass through the empty regions between masses, rather than
through a continuous energy density.  This means that the focusing
due to the notional fluid of an FRW cosmology is absent \cite{Bert,DR}.
Secondly, the redshift experienced by the photons is due to the
effect of anisotropic expansion integrated along their trajectories.
This does not, in general, correspond to the same redshift that would be experienced by
a photon in the `average' space-time.  Analytic expressions for these
quantities are given in paper I, together with the results of
numerical simulations.

In this paper we wish to include the effect of a Cosmological
Constant.  In some sense, this is a very straightforward extension of
the cosmology that was considered in paper I.  Crucially, however, it
will allow us to assess the influence of the effects we have uncovered on cosmological
parameter estimation, as well as provide us with a test of the validity
of our approach.  Huge resources are being
invested into gathering ever more observational data, and it is critical to the
success of the missions involved in this that we understand to the
highest degree possible the relationships between redshifts, luminosity
distances and the expansion of the Universe.  Any attempts at precision parameter
estimation \cite{prec}, or, for example, reconstructing the equation of state of
Dark Energy \cite{DE}, will be highly dependent on such considerations.

Let us consider a critical density cosmology where we have
a lattice of cells with  Schwarzschild-de Sitter geometry.  The
line-element inside each cell is then given by
\bea
\nonumber
ds^2 &=& -\left(1- \frac{2 m}{r}-\frac{\Lambda}{3}r^2 \right) d \tau^2
\\ && - 2
\sqrt{\frac{2 m}{r}+\frac{\Lambda}{3}r^2}
d\tau dr +dr^2 +r^2 d\Omega^2.
\label{ds}
\eea
Here the time coordinate $\tau$ is the proper time of a freely falling
observer with 4-velocity
\be
\label{u}
u^a=\left( 1; \sqrt{\frac{2 m}{r}+\frac{\Lambda r^2}{3}},0,0\right).
\ee
The relationship between $\tau$ and the Schwarzschild time coordinate
$t$ is given in Appendix A of paper I.  The reader is also referred to Section 3 of paper I for an
explanation of why $\tau$ corresponds to `cosmological time'.

Now, the principal difference between the Lindquist-Wheeler lattice construction in
General Relativity (GR) \cite{LW}, and Wigner-Seitz construction in
electromagnetism \cite{WS} (aside from the non-linearity of the field
equations in GR) is that in the case of the former the
lattice itself becomes dynamical.  In the case of a universe with
critical density, this leads to cosmological
evolution governed by the equation
\be
\label{H}
\frac{\dot{r}^2}{r^2} = \frac{2 m}{r^3} +\frac{\Lambda}{3},
\ee
where the over-dot here denotes differentiation with respect to $\tau$. If we
replace $r$ by the scale factor $a$, then this is clearly just the
Friedmann equation, with the usual solution
\be
\label{a}
a^3(\tau) = \frac{6 m}{\Lambda L_0^3} \sinh^2 \left[ \frac{\sqrt{3
      \Lambda}}{2} \tau \right],
\ee
where $L_0$ is the lattice spacing today (so that $a(\tau_0)=1$). The large-scale dynamics of
this model are therefore identical to an FRW model with $8 \pi
\rho=2m/(L_0a)^3$.  However, as discussed above, this does not necessarily mean that
the optical properties are also the same.

We know from our study of the $\Omega_{\Lambda} =0$ case that these corrections can
be considerable.  However, we also know that as
$\Omega_{\Lambda}\rightarrow 1$  they should vanish.  This is due
to the space-time approaching de Sitter space in this limit.  The
influence of the mass at the centre of each cell then becomes
negligible, as the space-time is dominated by $\Lambda$.  Of course,
the energy density associated with $\Lambda$ is constant everywhere,
and so we should expect any effects due to the discreteness of $m$ to
disappear.  Indeed, we find just such a convergence to occur in the
results below, and we consider this to be an important verification of the
model we are using (analogous to Shockley's `empty lattice test' with a
constant potential \cite{shock}, that is used to justify the
Wigner-Seitz approximation).

Now, to begin our study of redshifts and distance measures in this
space-time, let us consider a bundle of null rays that connect an
observer with an unobscured source.  The null geodesic equations  
can then be integrated to give
\bea
\label{B}
B &=& \left(1- \frac{2 m}{r} -\frac{\Lambda}{3}r^2 \right) \dot{\tau}
+\sqrt{\frac{2 m}{r} +\frac{\Lambda}{3}r^2}\dot{r}\\
\dot{r}^2 &=& B^2 -\frac{J^2}{r^2} \left(1- \frac{2 m}{r}
-\frac{\Lambda}{3}r^2 \right)\\
\dot{\theta}^2 &=& \frac{J^2}{r^4} - \frac{J^2_{\phi}}{r^4 \sin^2
  \theta}\\
\dot{\phi} &=& \frac{J_{\phi}}{r^2 \sin^2\theta},
\label{last}
\eea
where $B$, $J$ and $J_{\phi}$ are constants, and
over-dots here correspond to derivatives with respect to the affine
parameter, $\lambda$.  These
equations allow us to propagate null geodesics away from our observer,
and can be connected between lattice cells using the methods discussed
in paper I (with $2 m/r$ replaced by $2 m/r+\Lambda
r^2/3$, wherever it occurs in the matching conditions).

The redshift between emission and observation is given, as always,
by the ratio of frequencies at these two events.  For source
and observer both moving with 4-velocity $u^a$, as given in (\ref{u}), this is
\be
1+z = \frac{(-u^a k_a)\vert_e}{(-u^b k_b)\vert_o} =
\frac{\dot{\tau}\vert_e}{\dot{\tau}\vert_o},
\ee
where $k^a$ is the 4-vector tangent to the null geodesic, and
subscripts $e$ and $o$ denote quantities at emission and observation,
respectively.  The null geodesics equations, and redshift, can now be
calculated numerically.

For the $\Omega_{\Lambda}=0$ case we obtained in paper I the analytic approximation
\be
\label{z}
1+z \simeq (1+z_{FRW})^{\langle \gamma \rangle},
\ee
where $\langle \gamma \rangle =0.7$.  However, we know that in the limit
$\Omega_{\Lambda}\rightarrow 1$ that we should find $\langle
\gamma \rangle \rightarrow 1$, as the space-time approaches de Sitter
space.  We therefore expect that we should find $0.7 \lesssim \langle
\gamma \rangle \lesssim 1$, for $0 < \Omega_{\Lambda} < 1$.  

\begin{figure}[t]
\vspace{-5pt}
\center \epsfig{file=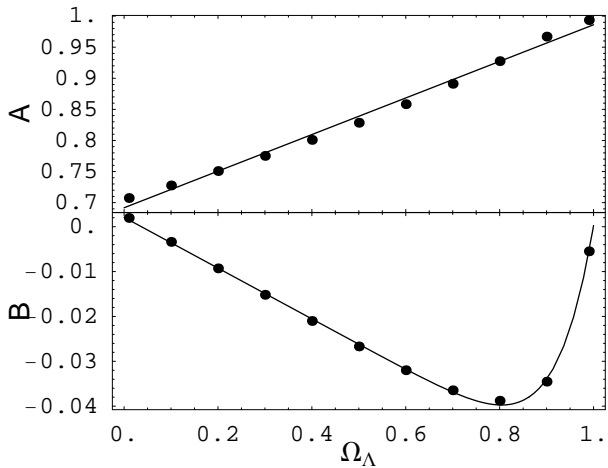,height=6.5cm} 
\vspace{-12pt}
\caption{Dots show the best fit values of $A$ and $B$ from
  Eq. (\ref{AandB}) for various values of $\Omega_{\Lambda}$.  The
  curves in the upper and lower plots are the fitted functions $0.69 +0.29
  \Omega_{\Lambda}$ and $0.0021 -0.057 \Omega_{\Lambda} +0.055
  \Omega_{\Lambda}^{12}$, respectively.}
\vspace{-12pt}
\label{AB}
\end{figure}

Our numerical simulations verify this expectation, and we find that
excellent fits can be achieved in the range $0<z<2$ by allowing $\langle \gamma \rangle$ to
be a running function of $z_{FRW}$, with the form
\be
\label{AandB}
\langle \gamma \rangle = A + B z_{FRW}.
\ee
Performing a least squares fit to the output of our numerical code we
find best fit values of $A$ and $B$ for various different values of
$\Omega_{\Lambda}$.  These are displayed in Fig. \ref{AB}, along with
the fitted functions $A=0.69 +0.29 \Omega_{\Lambda}$ and $B = 0.0021 -0.057
\Omega_{\Lambda} +0.055 \Omega_{\Lambda}^{12}$.  It can be seen that
$\langle \gamma \rangle$ approaches a constant as both
$\Omega_{\Lambda} \rightarrow 0$ (as found in paper I), and
as $\Omega_{\Lambda} \rightarrow 1$ (as the de Sitter limit is
approached).

To go further, and obtain expressions for angular diameter, and luminosity
distance in these models, we must integrate the Sachs optical
equations \cite{Sachs} along the geodesics that are solutions of
(\ref{B})-(\ref{last}).  Without rotation, these equations are
\bea
\label{sachs1}
\frac{d \tilde{\theta}}{d \lambda} +\tilde{\theta}^2+ \sigma^*
\sigma &=& -\frac{1}{2}R_{ab} k^a k^b\\
\label{sachs2}
\frac{d \sigma}{d \lambda} +2 \sigma \tilde{\theta} &=& C_{a b c d}
(t^*)^a k^b (t^*)^c k^d \equiv C,
\eea
where $\tilde{\theta}$ and $\sigma$ are the expansion
and complex shear scalars, respectively.  The $C_{a b c d}$
is Weyl's tensor, $R_{ab}$ is the Ricci tensor, and $t^a$ is a vector that is orthogonal to
$k^a$, null, and has a magnitude of $1$ (i.e. $t^a k_a=0$, $t^a t_a=0$
and $t^a (t^*)_a=1$). The initial conditions for integrating these equations are
$\sigma \vert_o =0$, $r_A \vert_o =0$ and $d r_A/d\lambda \vert_o
=\text{constant}$. 

%\begin{widetext}

\begin{figure*}[t]
\vspace{-4pt}
\center \epsfig{file=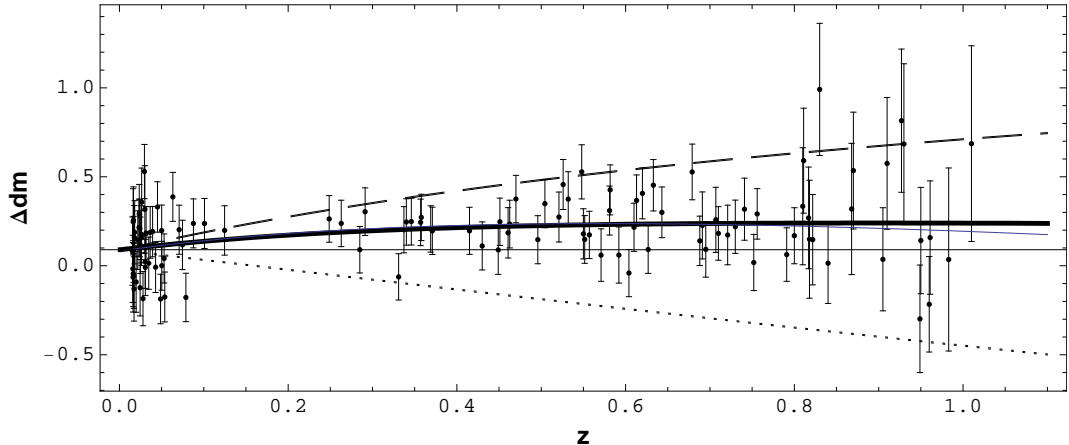,height=6cm}
\vspace{-10pt}
\caption{The best fit critical density models with discrete matter (thick
  line) and continuous fluid (thin line) sources.  The data is the
  115 high and low redshift supernovae from SNLS \cite{snls}, fitted
  to the discrete matter model with the SALT light curve fitter \cite{salt}.
  Einstein-de Sitter (dotted line) and de Sitter (dashed line) models
  are also shown, for reference.  The distance modulus, $\Delta$dm, is the magnitude
  of the source, minus the magnitude it would have at the same $z$ in
  an empty, open Milne universe.}
\vspace{-10pt}
\label{fig}
\end{figure*}

%\end{widetext}

Once the expansion scalar is known, then the angular diameter distance
is given by the integral
\be
\label{rA}
r_A \propto \text{exp}\left\{ \int_e^o \tilde{\theta} d \lambda \right\},
\ee
and the luminosity distance is given by Etherington's theorem
\cite{eth} as
$
%\label{Eth}
r_L = (1+z)^2 r_A.
$
For the geometry specified by (\ref{ds}) we then find that the driving
terms in (\ref{sachs1}) and (\ref{sachs2}) are given by 
\be
\label{Cphase}
R_{ab} k^a k^b =0 \;\;\;\; \text{and} \;\;\;\; C = \frac{3 m J^2}{r^5}  e^{i\Psi},
\ee
where $\Psi$ is a constant, specifying the complex phase.  These
equations are independent of $\Omega_{\Lambda}$, and so their solution
is very similar to the $\Omega_{\Lambda}=0$ case (up to the effect
$\Omega_{\Lambda}$ has on the trajectories $r=r(\lambda)$).

By numerically integrating (\ref{sachs1}) and (\ref{sachs2}) we find
that the shear is typically unimportant at
$z\lesssim 1$, unless a trajectory happens to pass very close to a
central mass.  At higher redshifts, however, the shear can accumulate and become more
important, eventually causing the divergence of the optical scalars
along some trajectories.  This corresponds to caustics occurring in the beams.

For the present purposes, at $z\lesssim 1$, it is sufficient to
neglect the shear, and set $\sigma=0$.  The solution to Eqs.
(\ref{sachs1}) and (\ref{rA}) is then given simply by $r_A \propto
\lambda$, and so we find that $r_L \propto (1+z)^2
\lambda$.  We now need to find the relationship between $\lambda$ and
$z$.  Using the expression for redshift, (\ref{z}), together with the
Friedmann equation, (\ref{H}), we can write
\bea
\label{z1}
1+z &=& \frac{\dot{\tau}_e}{\dot{\tau}_o} \simeq \left(
  \frac{a_o}{a_e} \right)^{\langle \gamma \rangle}\\
&=& \frac{1}{a_o \dot{\tau}_o} \frac{da_e}{d\lambda}
  \frac{1}{\sqrt{\Omega_m H_o^2 \frac{a_0}{a_e}+\Omega_{\Lambda} H_o^2
  \left( \frac{a_e}{a_o}\right)^2}}, \nonumber
\eea
where the usual expressions for $\Omega_m$ and
$\Omega_{\Lambda}$ have been used.  In general, this equation needs to
  be integrated numerically to find a solution for $\lambda (z)$.
  However, we find that by treating $\langle \gamma \rangle$ as a
  quasi-static variable we can obtain analytic results that are
  accurate to better than $1\%$.  In this case, the solution to Eq. (\ref{z1}) is
\be
\nonumber
\lambda \simeq -\frac{2}{\dot{\tau}_o H_o (3+2 \langle \gamma \rangle)
  \sqrt{\Omega_m}} \left[ \frac{f(z)}{(1+z)^{\frac{3+2\langle \gamma \rangle}{2\langle \gamma
    \rangle}}}-f(0) \right],
\ee
where
\be
\nonumber
f(z)= \;_2F_1 \left( \frac{3+2 \langle \gamma \rangle}{6},
\frac{1}{2}; \frac{9+2 \langle \gamma \rangle}{6}; -
\frac{\Omega_{\Lambda}}{\Omega_m} (1+z)^{-\frac{3}{\langle \gamma
    \rangle}} \right),
\ee
and$\;_2F_1(a,b;c;d)$ is the hypergeometric function \cite{hyper}.
We then have that, in the absence of shear, the luminosity distance is
well approximated as a function of $z$ by
\be
r_L \propto (1+z)^2 \left[ \frac{f(z)}{(1+z)^{\frac{3+2\langle \gamma \rangle}{2\langle \gamma
    \rangle}}}-f(0)\right].
\ee

We will now fit our model to the recent supernova
observations, and obtain an estimate for the difference in the best fit
value of $\Omega_{\Lambda}$ due to using a discretized matter
content, rather than a perfect fluid.
This will be done using the first-year Supernovae Legacy Survey (SNLS) data, consisting
of 115 supernovae \cite{snls}, calibrated with the Spectral Adaptive
Light-curve Template (SALT) fitter \cite{salt}.  Assuming a critical density, both FRW and our
model have 5 free parameters:  $\Omega_{\Lambda}$ and the 4 `nuisance'
parameters required to calibrate the supernova data.  These are the
absolute magnitude, the intrinsic error and the color and stretch
parameters used in the process of light curve fitting, $\{ M_0,
\sigma_{int}, \alpha, \beta \}$.

We show the best fits for both our model, and a spatially flat
continuous fluid FRW cosmology in Fig. \ref{fig}.  The fits are very
similar with a difference in log likelihood of just $\Delta \ln
\mathcal{L}=0.37$, in favor of the FRW model.  In fact, it can be seen from the
figure that the distance moduli of these two best fit models are
virtually identical, with the two solid lines only actually becoming
distinguishable by eye at $z\gtrsim 0.8$.  It seems clear that a lot more data
will be needed before there is any hope of distinguishing these two
curves observationally.

The interesting point, and the main result of this paper, is that
these two best fit models correspond to quite different values of
$\Omega_{\Lambda}$.  For the more usual perfect fluid FRW cosmology we recover
the standard result $\Omega_{\Lambda}^{FRW}= 0.74 \pm 0.04$.  For the
model with a matter content composed of discrete masses, however, we
obtain $\Omega_{\Lambda}=0.66 \pm 0.04$.  The best fit values of the
nuisance parameters are very similar in each case, as would be
expected from the similarity of the two distance moduli.

One may initially react by observing that this difference in
$\Omega_{\Lambda}$ is only at the level of 2$\sigma$, and as such
is not very significant.  Our point, however, is that irrespective of
how good the data eventually becomes, there will still be a difference
in $\Omega_{\Lambda}$ of $\sim 10\%$ between these two models.  This difference
is not due to any new physics, exotic matter content, or
unexpectedly large structures of voids in the universe, but only
to the fact that we have treated the matter content of the Universe in
our model as being discrete, rather than continuous.

Inevitably the reader will be concerned with the generality of this
result, and whether it should apply to the real Universe, or only to
our simple model.  We strongly suspect that the real Universe, with
its complicated networks of voids, filaments, walls and nodes \cite{fractal} will not behave
exactly as our model does, and that there will be new behaviour beyond that
which we have uncovered here.  In future work we hope to make progress
in generalizing our results to more realistic situations.  However, we
believe that there is good reason to expect some of the essential
features of our study to hold in the more general case.

Firstly, the photons we observe on Earth have not reached us by
travelling through a continuous fluid of critical density; they have
travelled through mostly empty space.  As such, they should not
experience the focusing that such a fluid would produce.  Photons
experience the geometry of space-time through which they have passed,
and {\it not} the global average.  Secondly, the expansion that
photons experience is also not the global average, it is the
integrated effect of the expansion which is strongly anisotropic at
any given point along their trajectory (as long as $\Omega_{m}>0$).  For further discussion of
these points, and discussion on the appropriateness of using an
approximate solution of Einstein's equations, we refer the reader to
paper I.

Of course, one will also be concerned about systematic errors, and the
validity of the assumptions that have gone into this model.  In future
publications we intend to establish the validity of these assumptions
through detailed investigation, but for now we already good reason to suspect that the approximations
that have been made are good.  In paper I it was shown that the
approximate nature of the boundary conditions between lattice cells is
insensitive to the details of the matching conditions, lending
credence to the idea of an average tangency between neighbouring space-like
surfaces.  Furthermore, we also
obtained in paper I analytic approximations that are in good agreement
with our numerical results.  These allow some physical insight into
the source of the effects we have uncovered, and further their
legitimacy.  Beyond this, the analogy with the highly successful
Wigner-Seitz construction and the recovery of the familiar optics of de
Sitter space in the appropriate limit is also encouraging.

In summary, we find that the fluid approximation in cosmology, while appearing
innocuous, can introduce considerable errors in interpreting
cosmological data.  Using a simple model of the Universe, with discrete
masses arranged on a regular lattice, we have shown that even if the
average dynamics of the Universe are unchanged, the
error introduced in the estimation of $\Omega_{\Lambda}$ due to
different optical properties can be of the
order of $10 \%$.  Such effects will need to be understood and accounted for
if we are to attempt precision cosmology.

%\newpage
% ------------------------ ACKNOWLEDGEMENTS ----------------------------------
\vspace{-20pt}
\section*{Acknowledgements}
\vspace{-15pt}

We are very grateful to George Ellis, Chris Clarkson and Lance Miller for suggestions and discussion.  TC
acknowledges the support of Jesus College and the BIPAC.

%\vspace{-20pt}

\end{document}